\documentclass[aps,prb,twocolumn,letterpaper]{revtex4}

\usepackage{amsmath,amssymb,amsfonts,amsthm}
\usepackage{graphicx,bm}

\begin{document}

\title{Mechanism of ferroelectric instabilities in non $d^0$
  perovskites: LaCrO$_3$ versus CaMnO$_3$}

\date{\today}

\author{Claude Ederer}
\email{edererc@tcd.ie}
\affiliation{School of Physics, Trinity College Dublin, Dublin 2, Ireland} 
\author{Tim Harris}
\affiliation{School of Physics, Trinity College Dublin, Dublin 2, Ireland} 
\author{Roman Kov\'a\v{c}ik}
\affiliation{School of Physics, Trinity College Dublin, Dublin 2, Ireland} 
 
\begin{abstract}
  The incompatibility of partial $d$ occupation on the perovskite
  $B$-site with the standard charge transfer mechanism for
  ferroelectricity has been a central paradigm in multiferroics
  research. Nevertheless, it was recently shown by density functional
  theory calculations that CaMnO$_3$ exhibits a polar instability that
  even dominates over the octahedral tilting for slightly enlarged
  unit cell volume. Here, we present similar calculations for
  LaCrO$_3$, which has the same $d^3$ $B$-site electron configuration
  as CaMnO$_3$. We find that LaCrO$_3$ exhibits a very similar, albeit
  much weaker, polar instability as CaMnO$_3$. In addition, while the
  Born effective charge (BEC) of the Mn$^{4+}$ cation in CaMnO$_3$ is
  highly anomalous, the BEC of Cr$^{3+}$ in LaCrO$_3$ is only slightly
  enhanced. By decomposing the BECs into contributions of individual
  Wannier functions we show that the ferroelectric instabilities in
  both systems can be understood in terms of charge transfer between
  TM $d$ and O $p$ states, analogously to the standard $d^0$
  perovskite ferroelectrics.
\end{abstract}

\pacs{}

\maketitle

\section{Introduction}
\label{sec:intro}

The relative scarcity of multiferroic materials has often been
explained by a chemical incompatibility between the factors that
promote ferroelectricity compared to the factors that promote magnetic
order.\cite{Hill:2000,Khomskii:2006} Even though several mechanisms
for ferroelectricity that are compatible with the simultaneous presence
of magnetic cations have been identified recently (see
e.g. Refs.~\onlinecite{Cheong/Mostovoy:2007,Khomskii:2009,Picozzi/Ederer:2009}),
the most common mechanism for driving polar displacements in typical
perovskite ferroelectrics seems to require a completely unoccupied $d$
shell of the transition metal (TM) cations on the perovskite
$B$-site. This statement is to some extent based on the observation,
that essentially all known perovskite ferroelectrics (e.g. BaTiO$_3$,
KNbO$_3$, PbTiO$_3$, Pb(Zr,Ti)O$_3$, etc.) contain formal $d^0$ TM
cations on the $B$-site. On the other hand a partial filling of the
electronic $d$ states is required in order to create a magnetic
moment,\cite{footnote:f-electron} and the resulting incompatibility
has been a central paradigm in multiferroics research over the past
few years.

The driving force behind the ferroelectric distortion in the $d^0$
perovskites, such as e.g. BaTiO$_3$ or KNbO$_3$, is related to
hybridisation between the filled oxygen 2$p$ states and the empty $d$
states of the TM
cation.\cite{Cohen:1992,Filippetti/Hill:2002,Khomskii:2006} A
ferroelectric displacement reduces the distance between the TM cation
and one or more of the surrounding oxygen anions, and thus strengthens
the corresponding covalent bonds, while slightly weakening the bonds
to the other surrounding oxygen anions where the corresponding bond
distance is increased. This leads to an overall gain in covalent bond
energy, which however is opposed by repulsive electro-static
forces. The gain in hybridisation energy is maximal if the antibonding
states with predominant TM $d$ character are completely empty, and is
zero if these antibonding states are completely filled.

Nevertheless, it has been shown recently that cubic perovskite
CaMnO$_3$ exhibits a similar ferroelectric instability which, even
though it is rather weak at ambient conditions, can be significantly
enhanced by applying negative
pressure.\cite{Bhattacharjee/Bousquet/Ghosez:2009} At the equilibrium
lattice constant the ferroelectric instability is suppressed by a much
stronger antiferrodistortive instability, i.e. a collective tilting of
the octahedral network, which leads to the observed $Pbnm$-distorted
perovskite structure of CaMnO$_3$. This antiferrodistortive mode is
rather independent of volume, so that the polar distortion becomes
dominant for slightly increased lattice constant. Very similar
behaviour has also been reported subsequently for SrMnO$_3$ and
BaMnO$_3$.\cite{Lee/Rabe:2010,Rondinelli/Eidelson/Spaldin:2009} These
calculations thus predict a new class of multiferroics, which could be
synthesised for example as epitaxial thin films, where strain can
stabilise the perovskite structure with enlarged lattice constant.
However, the question of what is the driving force behind the
ferroelectric instability in the corresponding systems, in particular
whether it is related to the above-described charge transfer
mechanism, has not been addressed in these previous studies.

Within a predominantly cubic crystal field, the $d^3$ case of
CaMnO$_3$ can also be interpreted as $e_g^0$ electron configuration,
analogously to the $d^0$ configuration found in most perovskite
ferroelectrics. In this case the empty $e_g$ states can in principle
provide the increase in bond energy required for the ferroelectric
instability, whereas the partial filling of the $t_{2g}$ states can
create a local magnetic moment. While the resulting gain in bond
energy is probably smaller than for the case of a $d^0$ TM cation, it
is not obvious that this gain is necessarily always smaller than the
opposing effect of the short range
repulsion.\cite{Khomskii:2006,Rondinelli/Eidelson/Spaldin:2009,Khomskii:2009}
In fact, exactly this scenario has already been discussed in
Ref.~\onlinecite{Khomskii:2006}, and arguments were given that in
addition to the usual repulsive forces, the Hund's rule coupling would
further disfavour the ferroelectric instability in these cases.


Here we address the question of what is the driving force behind the
ferroelectric instability in CaMnO$_3$ and related systems. We clarify
whether the ferroelectric distortion in these materials is driven by
charge transfer between the TM cation and its surrounding oxygen
anions, similar to the conventional $d^0$ perovskite
ferroelectrics. In order to investigate whether this effect is a
peculiarity that is specific to the Mn$^{4+}$ cation, or whether a
similar polar instability can also be observed in other magnetic
perovskites, we compare the case of CaMnO$_3$ with LaCrO$_3$. Under
equilibrium conditions LaCrO$_3$ exhibits a $Pbnm$-distorted
perovskite structure and G-type antiferromagnetic
order,\cite{Li_et_al:2002} similar to CaMnO$_3$. Furthermore, the
Cr$^{3+}$ cation on the perovskite $B$ site has a $d^3$ electron
configuration that is iso-electronic to the Mn$^{4+}$ cation in
CaMnO$_3$.

We calculate phonon frequencies and eigenmodes at the $\Gamma$-point
and at selected zone-boundary wave-vectors for LaCrO$_3$ in the ideal
perovskite structure at different volumes, and we then calculate and
compare Born effective charges for LaCrO$_3$, CaMnO$_3$, and the
nonmagnetic ferroelectric BaTiO$_3$. Furthermore, we analyse the
driving force for ferroelectric displacements in these three systems
by decomposing the Born effective charges into contributions of
individual Wannier functions.

\section{Computational details}

All results presented in this report are obtained using the Quantum
ESPRESSO package, employing a plane wave basis set and ultrasoft
pseudopotentials.\cite{Giannozzi_et_al:2009, Vanderbilt:1990} A
plane-wave kinetic energy cutoff of 35~Ry (420~Ry) is used for the
expansion of the wave-functions (charge density). The 3$s$ and 3$p$
semi-core states of Cr, Mn, and Ca, as well as the 5$s$ and 5$p$
states of La, are included in the valence. Different k-point meshes
are used to calculate the different quantities presented in this work,
with convergence verified in all cases.  Calculations are performed
using both the Generalised Gradient Approximation (GGA) and the
GGA+$U$ approach with $U_\text{eff}$ = 4~eV applied to the TM $d$
states.\cite{Perdew/Burke/Ernzerhof:1996,Dudarev_et_al:1998}

\section{Results}

\subsection{Phonon calculations}

Similar to Ref.~\onlinecite{Bhattacharjee/Bousquet/Ghosez:2009}, we
first identify potential phonon instabilities in cubic LaCrO$_3$ by
calculating eigenfrequencies of all zone-centre and various selected
zone-boundary modes at different lattice constants. We investigate
variations of $\pm 3$\% around an average lattice constant
$a=3.89$~\AA, which corresponds to the same volume per formula unit as
in the experimentally observed $Pbnm$ structure of
LaCrO$_3$.\cite{Li_et_al:2002} The relevant zone-boundary modes are
selected by decomposing the structural distortion between ideal
$Pm\bar{3}m$ symmetry and the experimental $Pbnm$ structure into
symmetry adapted modes, and then identifying the main components in
this decomposition. The ISODISPLACE utility is used for the mode
decomposition.\cite{Campbell_et_al:2006} The dynamical matrix is
obtained from the calculated forces created by small finite
displacements of the individual ions.

\begin{figure}
\includegraphics[width=0.7\columnwidth]{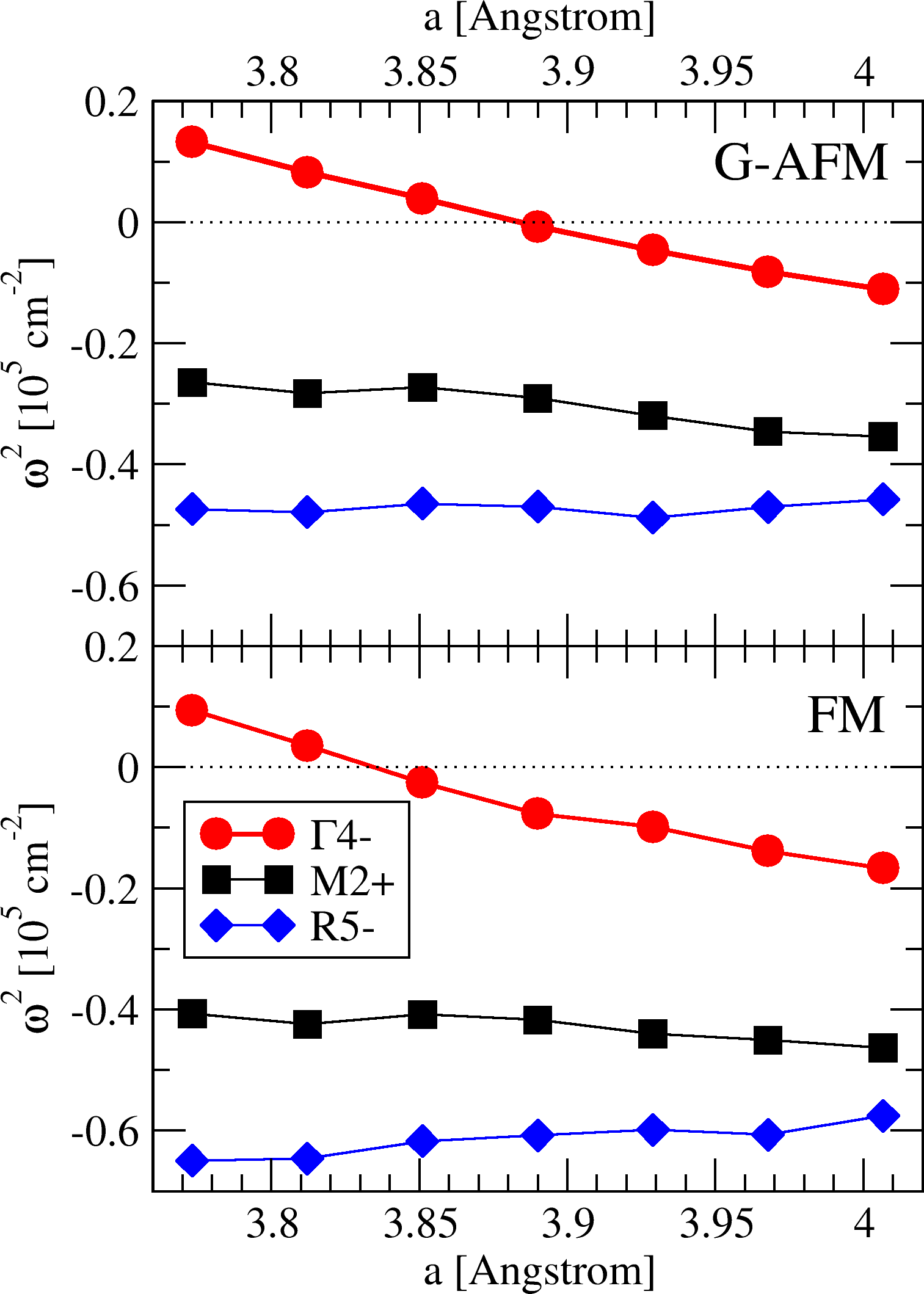}
\caption{Calculated eigenvalues $\omega^2$ of the dynamical matrix of
  cubic LaCrO$_3$ for selected modes as function of cubic lattice
  parameter $a$ using $U_\text{eff}=4$~eV. The upper panel corresponds
  to G-type antiferromagnetic order (G-AFM), the lower panel to the
  ferromagnetic case (FM). Negative $\omega^2$ indicates a structural
  instability.}
\label{fig:phonons}
\end{figure}

We find two strongly unstable antiferrodistortive zone-boundary modes
($R_5^-$ and $M_2^+$), which are responsible for the experimentally
observed $Pbnm$ ground state structure, and a soft polar
($\Gamma{}_4^-$) mode, which is unstable for larger lattice
constants.\cite{footnote:symmetry} The corresponding results are
presented in Fig.~\ref{fig:phonons} for both the ground state G-type
antiferromagnetic order and the ferromagnetic case. Similar to
CaMnO$_3$, the antiferrodistortive modes are rather insensitive to
volume, whereas the polar mode is strongly volume-dependent. It can
also be seen that ferromagnetic order leads to a further
destabilisation of all modes. This is a result of the fact that the
antiferromagnetic superexchange interaction is strongest for an ideal
180$^\circ$ TM-O-TM bond angle. In the magnetically unfavourable
ferromagnetic case the energy of the system can therefore be lowered
by distorting the bond angle away from 180$^\circ$.

It is clear from these results that qualitatively LaCrO$_3$ behaves
very similar to CaMnO$_3$. However, on a quantitative level, the polar
instability is significantly weaker in LaCrO$_3$ than in CaMnO$_3$
($\omega^2 = -0.11 \cdot 10^5$~cm$^{-2}$ for $+3$\% expansion in
lattice constant, compared to $\omega^2 = -0.36 \cdot
10^{5}$~cm$^{-2}$ for $+2$\% expansion in CaMnO$_3$). Even at rather
large volume, the polar instability in LaCrO$_3$ never dominates over
the antiferrodistortive modes.\cite{footnote:u}

\subsection{Born effective charges}

To further analyse the differences and similarities between LaCrO$_3$
and CaMnO$_3$, we calculate the Born effective charges (BECs) of the
$B$-site cations in both systems. The BEC tensor $Z^*_{i,\alpha\beta}$
describes the change in polarisation component $P_\alpha$ resulting
from a displacement of ion $i$ along Cartesian direction
$\beta$:\cite{Zhong/King-Smith/Vanderbilt:1994,Ghosez/Michenaud/Gonze:1998}
\begin{equation}
  Z^*_{i, \alpha \beta} = \frac{\Omega}{|e|} \frac{\partial P_\alpha}{\partial r_{i, \beta}} \quad .
\end{equation}
Here, $\Omega$ is the unit cell volume and $e$ is the electronic
charge. Anomalously enhanced BECs (compared to the formal charge
value) indicate that the corresponding displacements lead to strong
changes in hybridisation with the surrounding ions, and are generally
interpreted as ``smoking gun'' for the charge transfer mechanism
towards ferroelectricity described in
Sec.~\ref{sec:intro}.\cite{Zhong/King-Smith/Vanderbilt:1994,Ghosez/Michenaud/Gonze:1998}

\begin{table}
  \caption{Born effective charges $Z^*_{zz}$ of the TM cations in the three systems under consideration, calculated using the Berry phase approach, and compared to previously reported values and to the corresponding formal charges.}
\label{tab:BECs}
\begin{ruledtabular}
\begin{tabular}{lccc}
& BaTiO$_3$ & LaCrO$_3$ & CaMnO$_3$ \\
\hline
formal charge & 4 & 3 & 4 \\
$U_\text{eff} = 0$~eV & 7.31 & 3.62 & 7.61 \\
$U_\text{eff} = 4$~eV & --- & 3.58 & 7.66 \\
\hline
previous work ($U_\text{eff} = 0$~eV) & 7.25\cite{Ghosez/Michenaud/Gonze:1998} & 3.76\cite{Ray/Waghmare:2008} &  8.16\cite{Bhattacharjee/Bousquet/Ghosez:2009} \\
 & 6.89\cite{Filippetti/Hill:2002} & & 6.99\cite{Filippetti/Hill:2002}  
\end{tabular}
\end{ruledtabular}
\end{table}

We calculate BECs from the change in polarisation corresponding to
small finite displacements of the TM cations along the cubic axes. The
electric polarisation is calculated in two different ways: i) by using
the Berry phase approach,\cite{King-Smith/Vanderbilt:1993,Resta:1994}
and ii) by constructing maximally localised Wannier functions (MLWFs)
and monitoring changes in the centres of gravity of the individual
Wannier
orbitals.\cite{Marzari/Vanderbilt:1997,Vanderbilt/King-Smith:1993} The
BECs obtained using the Berry-phase approach are listed in
Table~\ref{tab:BECs}. In addition to the BECs of the Mn and Cr cations
in CaMnO$_3$ and LaCrO$_3$, respectively, the BEC for the Ti$^{4+}$
cation in the prototypical ferroelectric BaTiO$_3$ is also
calculated.\cite{footnote:becs}

It can be seen from Table~\ref{tab:BECs} that the effect of
$U_\text{eff}$ on the BECs is small and that all calculated values
agree well with previously reported data. Furthermore, in agreement
with Ref.~\onlinecite{Bhattacharjee/Bousquet/Ghosez:2009}, the BEC of
Mn$^{4+}$ in CaMnO$_3$ is very strongly enhanced compared to the
formal charge of $+4$. The anomalous part of the BEC amounts to about
90~\% of the formal charge (or $\sim$3.65 electrons), very similar to
the enhancement observed for the Ti$^{4+}$ cation in BaTiO$_3$. On the
other hand the BEC of Cr$^{3+}$ in LaCrO$_3$ is only increased by a
factor of 1.2 (or 0.6 electrons) compared to the formal charge of
$+3$. Thus, the differences in the BECs between LaCrO$_3$ and
CaMnO$_3$ seem consistent with the significantly weaker polar
instability of LaCrO$_3$ compared to CaMnO$_3$. However, we also point
out that there is no straightforward relationship between the
magnitude of the BECs and the tendency of a certain material towards
ferroelectricity (otherwise CaMnO$_3$ should be just as ferroelectric
as BaTiO$_3$).

\subsection{Wannier decomposition of BECs}

\begin{table}
  \caption{Decomposition of Born effective charges $Z^*_{zz}$ of the $B$ site cation in contributions of individual Wannier centres. Contributions of oxygen-centred MLWFs are averaged over both spin-projections, whereas TM $t_{2g}$ contributions correspond to the local majority spin-projection. All values are calculated for $U_\text{eff}=0$.}
\label{tab:MLWF-BECs}
\begin{ruledtabular}
\begin{tabular}{llccc}
& & BaTiO$_3$ & LaCrO$_3$ & CaMnO$_3$ \\
\hline
Mn/Cr & $d_{xy}(\uparrow)$ & --- & $-$1.081 & $-$1.019 \\
Mn/Cr & $d_{xz}$/$d_{yz}(\uparrow)$ & --- & $-$1.111 & $-$1.002 \\
\hline
O($z$) & $p_x$/$p_y$ & 0.853 & 0.239 & 0.516 \\
O($z$) & $p_z$ & 0.330 & 0.347 & 1.067 \\
O($z$) & $s$ & 0.267 & 0.212 & 0.331 \\
\hline
O($y$) & $p_x$ & $-$0.0812 & $-$0.059 & $-$0.088 \\
O($y$) & $p_y$ & $-$0.141 & $-$0.217 & $-$0.300 \\
O($y$) & $p_z$ & $-$ 0.092 & 0.023 & 0.165 \\
O($y$) & $s$ & $-$0.072 & $-$0.129 & $-$0.124\\
\hline
\multicolumn{2}{l}{total valence} & 3.065 & $-$2.760 & 0.451 \\
\multicolumn{2}{l}{semi-core} & $-$7.716 & $-$7.628 & $-$7.877 \\
\multicolumn{2}{l}{ionic core} & 12.000 & 14.000 & 15.000 \\
\hline
\multicolumn{2}{l}{total BEC} & 7.349 & 3.612 & 7.573 \\
\multicolumn{2}{l}{formal} & 4 & 3 & 4 \\
\end{tabular}
\end{ruledtabular}
\end{table}

To further analyse the origin of the enhanced BECs in the three
investigated systems, we decompose the electric polarisation
calculated via MLWFs into contributions of the individual Wannier
orbitals $|w_n\rangle$:\cite{Vanderbilt/King-Smith:1993}
\begin{equation}
  P = P_\text{ionic core} - \frac{|e|}{\Omega} \sum_n \langle w_n | r | w_n \rangle \quad .  
\end{equation}
This translates into a corresponding decomposition of the
BECs:
\begin{equation}
  Z^* = Z_\text{ionic core} + \sum_n Z^*_n \quad ,
\end{equation}
with:
\begin{equation}
  Z^*_n = - \frac{\Delta r_n}{\Delta r_\text{ion}}
\end{equation}
Here, $\Delta r_n = \langle w_n(\Delta r_\text{ion}) | r | w_n(\Delta
r_\text{ion}) \rangle - \langle w_n(0) | r | w_n(0) \rangle$ is the
displacement of the Wannier centre $n$ resulting from the ionic
displacement $\Delta r_\text{ion}$. 

\begin{figure}
  \includegraphics[width=0.5\columnwidth]{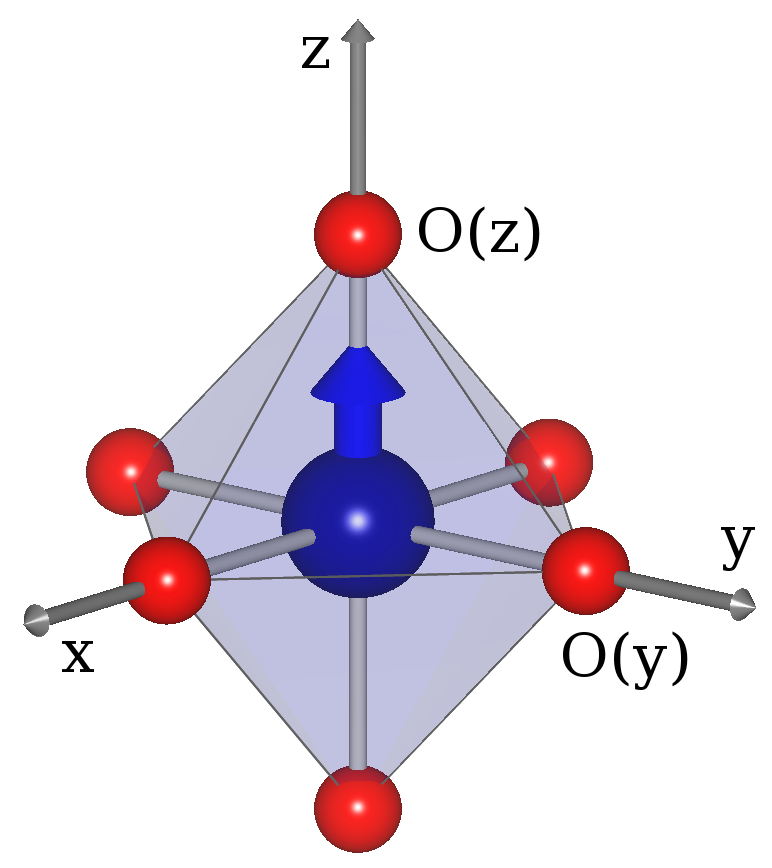}
  \caption{Geometry for the calculation of BECs listed in
    Table~\ref{tab:MLWF-BECs}. O($y$) and O($z$) are the oxygen anions
    situated adjacent to the displaced TM cation along the $y$ and $z$
    directions, respectively. The thick arrow indicates the
    displacement direction ($z$ direction). Picture generated using
    VESTA.\cite{vesta}}
\label{fig:geometry}
\end{figure}

\begin{figure}
\includegraphics[width=\columnwidth]{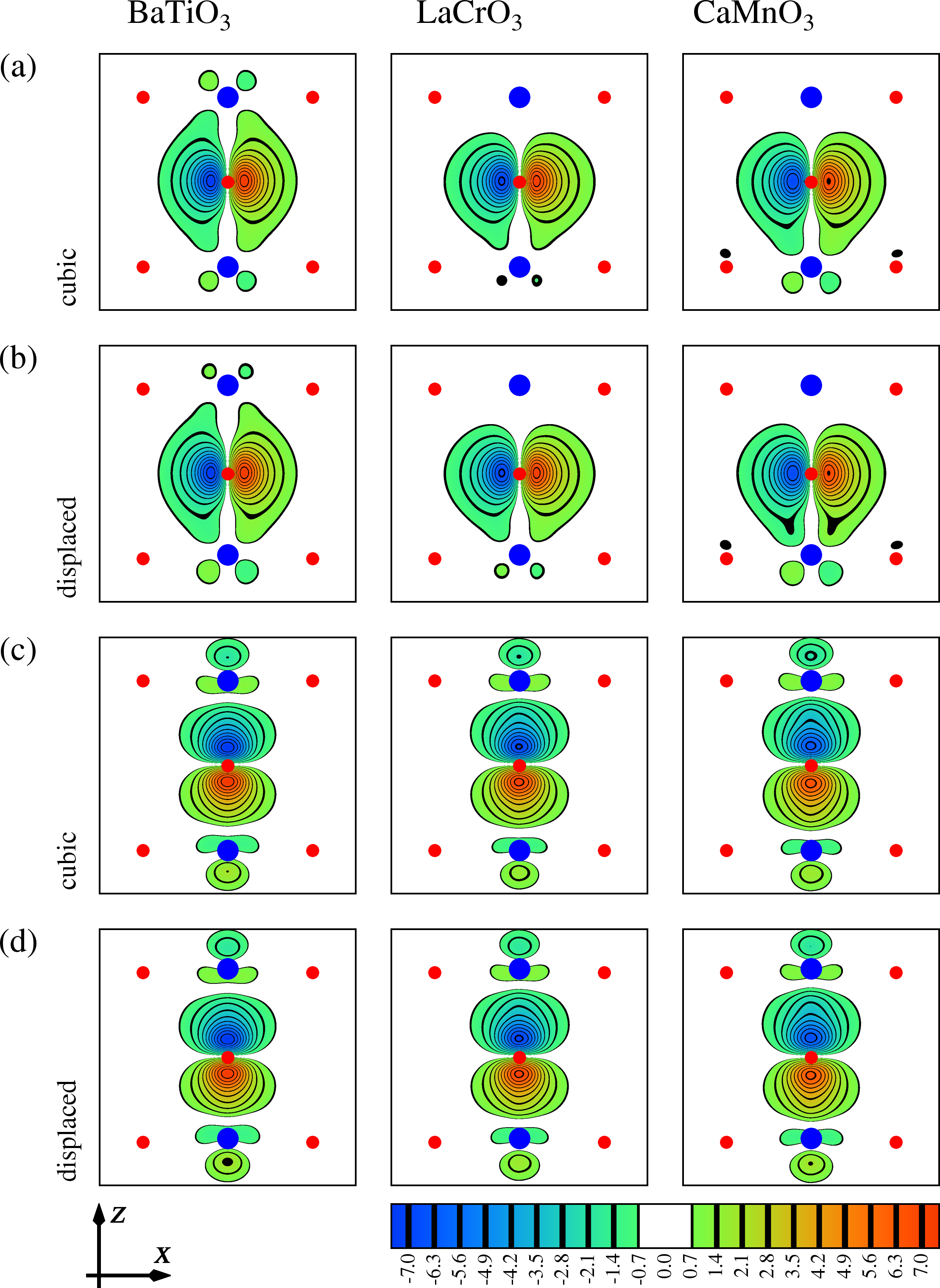}
\caption{(Color online) Maximally localised Wannier functions
  corresponding to $\pi$-oriented $p_x$ orbitals ((a), (b)) and
  $\sigma$-oriented $p_z$ orbitals ((c), (d)) centred on O($z$) for
  BaTiO$_3$ (left column), LaCrO$_3$ (middle column) and CaMnO$_3$
  (right column). (a) and (c) correspond to the cubic structures,
  whereas in (b) and (d) the TM cations were displaced along $+z$ by
  1\% of the cubic lattice constants. Shown are cuts through the
  $x$-$z$ plane. The positions of the oxygen anions/TM cations are
  indicated by the filled small/large (red/blue) circles.}
\label{fig:MLWFs}
\end{figure}

For each of the three systems under consideration, the occupied
valence and semi-core states form energetically isolated groups of
bands with a specific dominant atomic and orbital character: TM
semi-core 3$s$ and 3$p$, $A$-site cation semi-core $s$ and $p$, and
oxygen 2$s$ and 2$p$. In the case of LaCrO$_3$ and CaMnO$_3$, the
oxygen 2$p$ bands are inter-mixed with the local majority spin TM
$t_{2g}$ states. We construct separate MLWFs for each isolated set of
occupied bands using the Wannier90 code.\cite{Mostofi_et_al:2008} The
resulting MLWFs exhibit a clear atomic and angular momentum ($lm$)
character, albeit with a certain amount of admixture of other atomic
orbitals on the surrounding ions due to hybridisation (see
Fig.~\ref{fig:MLWFs} for some examples). We use this atomic and
orbital character to label each MLWF. The individual contributions to
the TM cation BECs of all individual valence MLWFs as well as the sum
over all semi-core contributions are listed in
Table~\ref{tab:MLWF-BECs}. O($z$) and O($y$) denote the two symmetry
inequivalent oxygen anions situated along and perpendicular to the
displacement direction of the TM cation (see Fig.~\ref{fig:geometry}).
It can be seen that the values for the total BECs calculated via MLWFs
which are listed in Table~\ref{tab:MLWF-BECs} are in good agreement
with the values calculated via the Berry-phase approach listed in
Table~\ref{tab:BECs}. The individual contributions to the BECs for
BaTiO$_3$ are also in good agreement with results of a similar
decomposition presented in Ref.~\onlinecite{Marzari/Vanderbilt:1998}.

Within a completely ionic picture, i.e. without inter-site
hybridisation, the contributions to the BECs from the occupied TM
$t_{2g}$ orbitals in LaCrO$_3$/CaMnO$_3$ would be exactly equal to
$-1$, whereas the contributions from all oxygen orbitals would be
identically zero. In addition, with the core/valence separation used
in our pseudopotentials, the total contribution of all semi-core
states would be exactly equal to $-8$ for all three systems. The
$t_{2g}$ contributions to the BECs in both LaCrO$_3$ and CaMnO$_3$ are
indeed very close to the nominal value of $-1$, and the semi-core
contributions to $Z^*$ are also close to $-8$, deviating only by about
0.1-0.4 electrons from this value. In contrast, strong anomalous
contributions are found for the oxygen $s$ and $p$ orbitals, with the
largest contributions resulting from the $p$ orbitals corresponding to
the oxygen anions O($z$) situated above and below the $B$-site cation
along the displacement direction ($z$-direction, see
Fig.~\ref{fig:geometry}). 

The large positive contributions of these orbitals indicate that the
centres of the corresponding MLWFs shift towards the TM cation that is
moved closer to the oxygen. This represents a net electron flow from
the oxygen anion towards the TM cation. Fig.~\ref{fig:MLWFs} depicts
the changes in the O($z$) $p$ MLWFs for a displacement of the TM
cation along $+z$ by 1\% of the cubic perovskite lattice constant. We
point out that this displacement is much larger than the one used for
the calculation of the BECs (which was only $\sim$0.2~\%). A larger
displacement was used in Fig.~\ref{fig:MLWFs} to make the changes in
the MLWFs more visible. It can be seen that the displacement of the TM
cation towards the oxygen anion increases the amplitude of the
corresponding atomic $d$ character in the O($z$) $p$ MLWFs compared to
the cubic case. This change in hybridisation shifts the centre of the
Wannier function towards the closer TM cation, leading to the
anomalous contribution to the BECs.

In BaTiO$_3$, where the Ti $d$ states are empty and therefore do not
contribute to the polarisation, the anomalous contribution to the BECs
listed in Table~\ref{tab:MLWF-BECs} are strongest for the $\pi$-type
($p_x$/$p_y$) orbitals, i.e. for the $p$ orbitals that are oriented
perpendicular to the Ti-O bond and hybridise with the empty $t_{2g}$
states on the Ti cation. The contribution of the $\sigma$-type ($p_z$)
orbital, which hybridises with the Ti $e_g$ states, is more than a
factor of two smaller.

If we compare this to the case of LaCrO$_3$, where the majority spin
$t_{2g}$ states are filled, the relative contributions of the two
types of $p$ states change significantly. The contribution of the
$p_z$ orbital is very similar to the case of BaTiO$_3$, but the
contributions of the $p_x$ and $p_y$ orbitals are strongly reduced and
are now smaller than the $p_z$ contribution.

This effect is related to a strong reduction of the $d$-like ``tails''
in the O($z$) $p_x$ and $p_y$ MLWFs in LaCrO$_3$ compared to
BaTiO$_3$, which can be seen from Fig.~\ref{fig:MLWFs}a and
Fig.~\ref{fig:MLWFs}b. In BaTiO$_3$, these tails represent the TM $d$
character contained in the nominal oxygen $p$ bands, which, as
discussed in the introductory paragraphs, are \emph{bonding} states
that result from hybridisation between atomic TM $d$ and oxygen $p$
orbitals. Once the corresponding \emph{antibonding} states (or parts
thereof) become occupied, the different orbital contributions can be
separated into different MLWFs with appropriate orbital character.

More specifically, we note that the MLWFs shown in
Fig.~\ref{fig:MLWFs} correspond to the global ``spin-up''
projection. For BaTiO$_3$ the two spin-projections are of course
identical, whereas for the magnetic systems the ``spin-up'' and
``spin-down'' MLWFs centred at O($z$) are related to each other by
space inversion. For the specific cases shown in Fig.~\ref{fig:MLWFs}
the local magnetic moment of the Cr/Mn cation located at
$+\tfrac{a}{2}\hat{z}$ relative to the central oxygen anion is
parallel to the global spin-up direction, whereas the local magnetic
moment of the Cr/Mn cation at $-\tfrac{a}{2}\hat{z}$ is parallel to
the global spin-down direction. It can be seen that the $t_{2g}$-tails
of the $p_x$ MLWFs at $+\tfrac{a}{2}\hat{z}$ have completely
vanished. This is due to the effect described above, i.e. the
corresponding orbital character has been transferred to the
$t_{2g}(\uparrow)$ MLWFs. In addition, the $t_{2g}$-tail at
$-\tfrac{a}{2}\hat{z}$ is also significantly reduced, because the
global spin-up direction at this TM site correspond to the local
minority spin character. The corresponding $t_{2g}$ states are
therefore higher in energy, which reduces the amount of $p$--$t_{2g}$
hybridisation at this site. As a result, the $p_x$/$p_y$ MLWFs in
LaCrO$_3$ resemble more closely the corresponding atomic orbitals
compared to the $p_x$/$p_y$ MLWFs in BaTiO$_3$, and the contributions
to the BECs become less anomalous.

The same hierarchy between the $p_x$/$p_y$ and $p_z$ contributions as
in LaCrO$_3$ can be observed for CaMnO$_3$, but the overall magnitude
of both contributions is strongly enhanced in the latter. The
contribution of the O($z$) $p_z$ MLWF in CaMnO$_3$ is even three times
larger than the corresponding contribution in BaTiO$_3$.

\subsection{Densities of states}

\begin{figure}
\includegraphics[width=0.7\columnwidth]{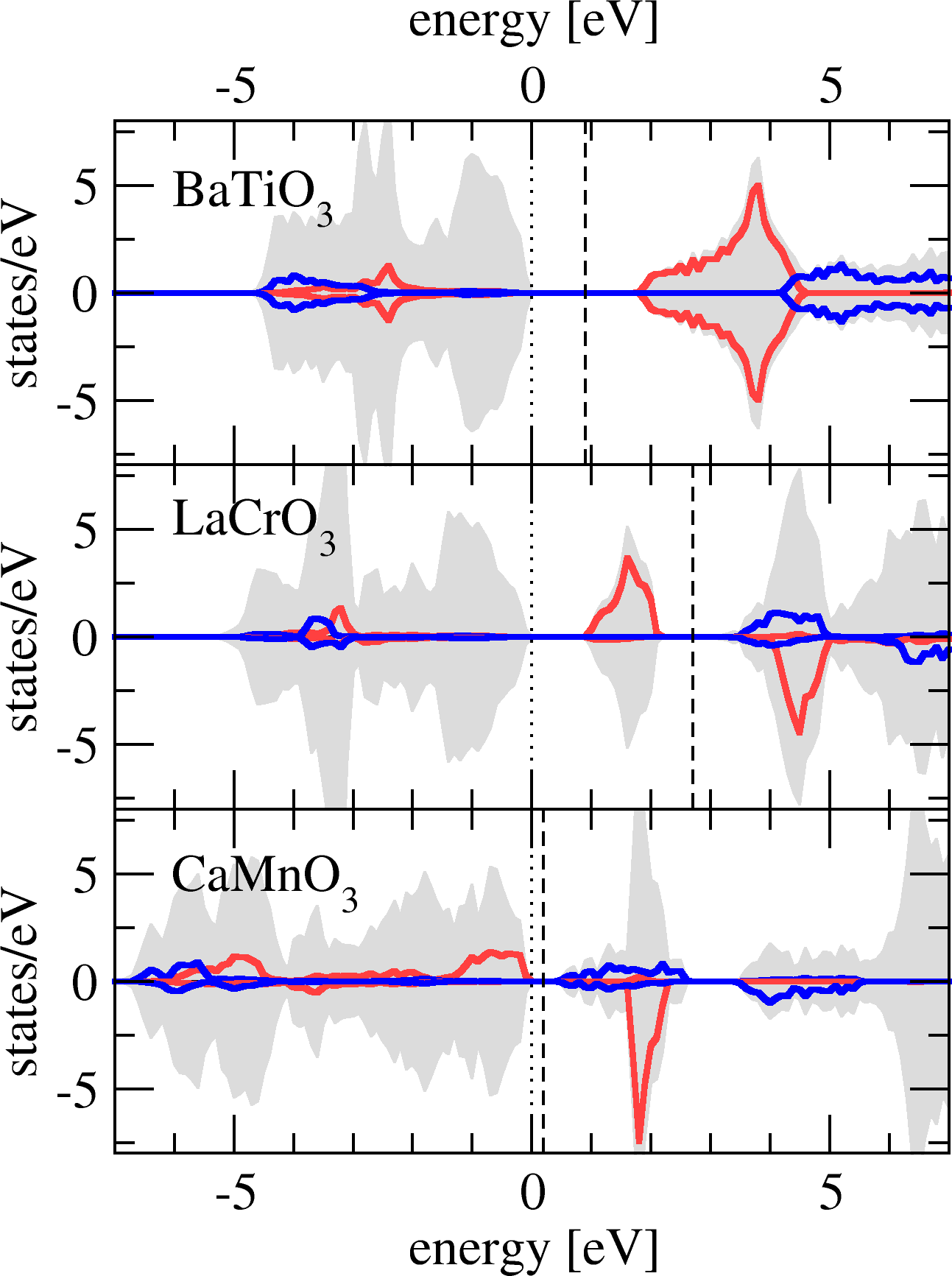}
\caption{(Color online) Spin-resolved total and projected DOS fur
  cubic BaTiO$_3$ (top), LaCrO$_3$ (middle), and CaMnO$_3$ (bottom),
  calculated within GGA. The shaded curves represent the total DOS,
  the bright/dark (red/blue) lines represent the $t_{2g}$/$e_g$
  projected DOS of the $B$-site cations. Different spin projections
  are shown with different sign. For better comparison, the DOS are
  aligned to the top of the oxygen $p$ bands. Mid-gap (Fermi) levels
  are indicated by dashed vertical lines.}
\label{fig:dos}
\end{figure}

The origin of the strong enhancement of the individual contributions
to the BECs in CaMnO$_3$ can be rationalised by looking at the
differences in the electronic structure of the three materials. The
total and projected densities of states (DOS) for BaTiO$_3$
(nonmagnetic), LaCrO$_3$, and CaMnO$_3$ (both G-type AFM) are shown in
Fig.~\ref{fig:dos}. It can be seen that in BaTiO$_3$ and LaCrO$_3$ the
TM $d$ bands are situated well above the oxygen $p$ bands. In
particular the bottom of the TM $e_g$ states is about 3.5-4~eV above
the top of the predominantly oxygen $p$ bands in both systems. On the
other hand in CaMnO$_3$ the corresponding energy separation is much
smaller, with the bottom of the Mn $e_g$ states only about 0.5~eV
above the oxygen $p$ bands. This small energy separation leads to very
strong hybridisation between oxygen $p$ and TM $d$ levels in
CaMnO$_3$. The much stronger hybridisation in CaMnO$_3$ compared to
both LaCrO$_3$ and BaTiO$_3$ will thus enhance all resulting effects,
including the anomalous contributions to the BECs. The strong
hybridisation in CaMnO$_3$ therefore compensates the reduced
$p_x$/$p_y$ contribution relative to $p_z$, leading to the large total
BEC (of the same magnitude as in BaTiO$_3$) and the rather strong
polar instability compared to LaCrO$_3$.

As a final technical note, we point put that the weak effect of
$U_\text{eff}$ on the BECs in CaMnO$_3$ and LaCrO$_3$ is a result of
the specific projection on (non-orthogonal) atomic orbitals used for
the LSDA+$U$ implementation within Quantum ESPRESSO. For both
LaCrO$_3$ and CaMnO$_3$ the resulting orbital occupation matrix
element corresponding to the majority spin $e_g$ states is
approximately equal to 0.5. Since the LSDA+$U$ potential shift of the
TM $d$ states is given by $\Delta V_m^\sigma =
\tfrac{U_\text{eff}}{2}(\tfrac{1}{2}-n_m^\sigma)$,\cite{Dudarev_et_al:1998}
the energy separation between the $e_g$ states and the oxygen $p$
bands is thus not affected by the value of $U_\text{eff}$ for
$n_{e_g}^\uparrow \approx 0.5$. According to the preceding discussion
it is essentially this energy difference that determines the magnitude
of the anomalous contribution to the TM BECs in LaCrO$_3$ and
CaMnO$_3$, which are therefore rather independent of $U_\text{eff}$.

\section{Summary and Conclusions}

In summary, we have shown that LaCrO$_3$ exhibits a strongly
volume-dependent polar instability very similar to CaMnO$_3$. However
the polar instability is significantly weaker in LaCrO$_3$ than in
CaMnO$_3$. This is consistent with the much less anomalous BEC of the
Cr$^{3+}$ cation in LaCrO$_3$ compared to the strongly enhanced BEC of
Mn$^{4+}$ in CaMnO$_3$. By decomposing the BECs in contributions of
individual Wannier functions, we could show that in both cases the
largest anomalous contributions stem from $\sigma$-oriented O $p$
states of the oxygen anions adjacent to the TM cation along the
displacement direction, whereas the corresponding $\pi$ contributions
are reduced compared to the $d^0$ ferroelectric BaTiO$_3$. This
reduction of the $\pi$-type contributions to the BECs relative to the
$\sigma$-type contribution is due to the occupation of the majority
spin $t_{2g}$ states in LaCrO$_3$ and CaMnO$_3$. The strong overall
enhancement of the anomalous charges in CaMnO$_3$ can be explained by
the near degeneracy of O $p$ and TM $d$ states in CaMnO$_3$, which
leads to very strong hybridisation and thus enhances all related
effects.  The polar instabilities in both magnetic perovskites can
therefore be understood in terms of charge transfer between TM $d$ and
O $p$ states, analogously to the standard $d^0$ perovskite
ferroelectrics.

\begin{acknowledgments}
  This work was supported by Science Foundation Ireland under grant
  numbers SFI-07/YI2/I1051 (PIYRA) and SFI-UR-I1531 (SURE UREKA site),
  and made use of computational facilities provided by the Trinity
  Centre for High Performance Computing.
\end{acknowledgments}

\bibliography{references}

\begin{thebibliography}{30}
\bibitem[{foo({\natexlab{a}})}]{footnote:f-electron}
\bibinfo{note}{In the present discussion we do not consider $f$-electron
  magnetism.}
\bibitem[{foo({\natexlab{c}})}]{footnote:u}
\bibinfo{note}{We have verified that the calculated phonon eigenvalues do not
  depend strongly on the specific value of $U_\text{eff}$ used in the
  calculation.}
\end{thebibliography}

\end{document}